# Numerical Analysis of Photon Absorption of Gate-defined Quantum Dots Embedded in Asymmetric Bull's-eye Optical Cavities

SANGMIN JI[1, 2 *] AND SATOSHI IWAMOTO[1, 2]

[1]*Institute of Industrial Science, The University of Tokyo, 4-6-1 Komaba Meguro, Tokyo, 1530041, Japan*
[2]*Research Center for Advanced Science and Technology, The University of Tokyo, 4-6-1 Komaba Meguro, Tokyo, 1530041, Japan*
*\*smji@iis.u-tokyo.ac.jp*

**Abstract:** Improving the photon–spin conversion efficiency without polarization dependence is a major challenge in realizing quantum interfaces gate-defined quantum dots (QDs) for polarization-encoded photonic quantum network systems. Previously, we reported the design of an air-bridge bull's-eye cavity that enhances the photon absorption efficiency of an embedded gate-defined QD regardless of the photon polarization. Here, we numerically demonstrate that a further 1.6 times improvement in efficiency is possible by simply adjusting the distance of the substrate from the semiconductor slab where the bull's-eye structure is formed. Our analysis clarifies that the upward-preferred coupling and narrow far-field emission pattern realized by substrate-induced asymmetry enable the improvement.

## 1. Introduction

A quantum network [1,2] providing coherent links of distant stationary qubits via flying qubits can advance quantum informatic apparatuses for communications [3], computing [4], and sensing [5]. Photonic quantum interfaces [6] that convert quantum states between stationary matter qubits and flying photonic qubits are essential building blocks in such quantum networks. Several platforms are candidates for hosting stationary qubits that enable photonic quantum interfaces, including superconducting circuits [7], Rydberg atoms [8], trapped ions [9], color centers in diamonds [10,11], and semiconductor quantum dots (QDs) [12,13]. Gate-defined QDs [14–16], which are formed by laterally confining the carriers in semiconductor quantum wells (QWs) with an applied gate voltage, are also attractive solid-state hosts for stationary spin qubits. Using spins in Si-based gate-defined QDs [17], two-qubit gate operations with high fidelity above the error correction threshold have been demonstrated recently [18,19]. These Si QDs can be realized with complementary-metal–oxide–semiconductor-compatible fabrication processes [20], which is a powerful advantage, particularly for the scalable integration of qubits. As for implementing photonic quantum interfaces, GaAs-based gate-defined QDs are suitable. Utilizing the polarization-dependent optical absorption in III-V semiconductor QWs [21,22], angular momentum transfer from a photon to an electron in a gate-defined GaAs QD has been experimentally realized [23,24]. However, the photon-to-spin conversion efficiency is currently low (~$10^{-5}$ [25]) because of the limited photon absorption efficiency of the QDs, where the interaction volume with incident photons is small. This efficiency problem is a major bottleneck for realizing quantum repeaters that exploit photon-to-spin conversion and becomes even more significant for entanglement distribution requiring the use of multiple repeaters [26].

A promising approach for the improving efficiency is to embed QDs in optical cavities. Because polarization–spin correspondence is essential for polarization-encoded qubits, the optical cavity must be polarization-independent. Although using time-bin qubits [27] may alleviate this polarization independence, the conservation of polarization information provides additional degrees of freedom in quantum network configurations. The so-called bull's-eye structure [28–32] can exhibit polarization-independent features because of its centrosymmetric nature. Previously, we designed an air-bridge bull's-eye optical cavity and numerically demonstrated that it could improve the photon absorption efficiency of an embedded gate-defined QD by a factor of more than 400 [33]. Recently, we reported the fabrication and preliminary proof-of-principle experiments demonstrating polarization-independent photon absorption enhancement using the designed air-bridge bull's-eye cavities [34].

In this study, we discuss the possibility of further absorption enhancement due to directional coupling induced by out-of-plane asymmetry in the cavity. The coupling rates for the out-of-plane directions are unbalanced in such an asymmetric cavity; therefore, enhanced or suppressed coupling rates can be obtained in the desired direction. Several methods, such as controlling etching depth [28,35] and employing one-side reflectors, including distributed Bragg reflectors [36] or buried metal layers [31,37,38], have been reported for realizing optical cavities with out-of-plane asymmetry. However, these methods may limit productivity because of the requirement for precise control of the fabrication process or for a wafer with a complicated layer structure. Here, we numerically investigate the effect of asymmetry induced by the presence of substrate underneath the bull's-eye structure; the substrate-induced asymmetry can be easily realized in real device structures where the device is suspended in the air by removing a sacrificial layer on the substrate. Although the effect of the substrate on the radiation properties of photonic crystal cavities has been discussed in a previous study [39], we discuss its application in the case of cavity-enhanced absorption. Compared to the free-standing air-bridge bull's-eye cavity, the absorption efficiency of the gate-defined QDs in the asymmetric bull's-eye cavity can be enhanced by about 1.6 times by adjusting the air-gap distance between the substrate and the slab where the bull's-eye structure is formed. The theoretical analysis based on the equations representing the cavity-enhanced QD light absorption derived in our previous study [33] shows that the directionality realized by the asymmetry serves as a key factor for increased efficiency. This result suggests that utilizing substrate-induced asymmetry is a promising approach for realizing efficient photon–spin interfaces using gate-defined QDs.

## 2. Model and key factors for absorption enhancement

### 2.1. Cavity structure and mode characteristics

The cavity structure we considered is schematically shown in Fig. 1(a). A bull's eye structure with metal electrodes for creating a gate-defined QD is formed in a 165-nm-thick semiconductor slab ($T_{\text{slab}} = 165$ nm) comprising a 15-nm-thick GaAs QW ($T_{\text{QW}} = 15$ nm, $z = 0$ at the center of the QW layer). A GaAs substrate, which is $T_{\text{gap}}$ apart from the slab, is included in the model. The presence of the substrate causes the cavity to be vertically asymmetric. We used the three-dimensional finite difference time domain (3D-FDTD) method to calculate the light absorption efficiency of the QD (denoted as $\eta_{\text{abs}}^{\text{QD}}$) and related features (field distributions and quality factors of the cavity modes). For the calculation of absorption efficiency, a vertically incident Gaussian beam (beam waist diameter $w = 1$ $\mu$m) was used as the excitation source. The gate-defined QD was modeled as a cuboid with a lateral-size $S_{\text{QD}}$ of 50 nm × 50 nm and a thickness of $T_{\text{QW}}$.

As the bull's-eye structure, we used our previously designed air-bridge bull's-eye cavity [33] (Fig. 1(b)). The bull's-eye structure comprised a six-period circular Bragg grating (period $a = 340$ nm, air-gap filling factor $G = 0.4$) surrounding a circular cavity with diameter $A_c = a$. Two straight channels (width $W = 100$ nm) crossing the center provided space for gate formation and enabled the air-bridge structure [34]. The channel width was selected to provide sufficient space for electrode formation [40] while keeping a reasonable absorption enhancement [33]. The electrodes consisted of four gold metal lines, which were 36 nm wide and 40 nm thick. The $C_{4v}$ symmetry of the entire structure guarantees that the cavity supports doubly degenerate modes, which are essential for achieving a polarization-independent response. Figure 1(c) shows the field distribution of the doubly degenerate dipole-like modes supported in bull's-eye structure without the substrate. The resonant wavelength $\lambda_0$ of the degenerate modes was 806 nm. The employed second-order Bragg condition ($a = \lambda_0/n_{\text{eff}}$, where $n_{\text{eff}} \cong G + (1-G)n_{\text{slab}}$, $n_{\text{slab}} = \frac{T_{\text{QW}}}{T_{\text{slab}}}n_{\text{GaAs}} + \frac{T_{\text{slab}}-T_{\text{QW}}}{T_{\text{slab}}}n_{\text{AlGaAs}}$, $n_{\text{GaAs}} = 3.659$, and $n_{\text{AlGaAs}} = 3.346$) causes the radiation pattern of the mode to be concentrated in the vertical direction, resulting in a larger far-field overlap with the Gaussian excitation beam. The degenerate modes overlap at the cavity center; thus, the optical absorption of the QD (shown as green squares in Fig. 1(c)) is polarization-independent.

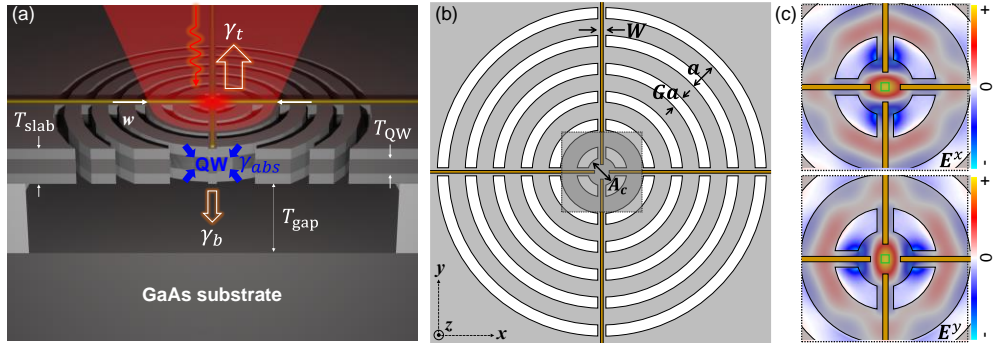

Fig. 1. (a) Schematic of vertically asymmetric air-bridge bull's-eye cavity. GaAs substrate is $T_{\text{gap}}$ apart from the cavity. The decay rates of the cavity mode due to the top-emitted, bottom-emitted, and QW absorption are denoted by $\gamma_{\text{t}}$, $\gamma_{\text{b}}$, and $\gamma_{\text{abs}}$, respectively. An excitation beam with a Gaussian distribution is focused at the center of the QW layer ($z = 0$) with a beam waist of $w$. (b) Schematic of bull's-eye optical cavity used in this study. (c) Distributions of main electric field of doubly degenerate modes supported in the cavity. The green squares at the cavity center represent the gate-defined QD formed by the electrodes.

### 2.2. Analytical expression for cavity-enhanced light absorption efficiency

In our previous study, we derived an analytical expression for the cavity-enhanced light absorption efficiency of a gate-defined QD using the temporal coupled-mode theory. Following the result, the absorption efficiency of the QD in a cavity at the resonant wavelength of $\lambda_0$ is obtained as [33]

$$\eta_{\text{abs}}^{\text{QD}}(\lambda_0) = \frac{S_{\text{QD}}}{S_{\text{eff}}}\eta_{\text{far}}M. \quad (1)$$

$S_{\text{eff}}$ represents an effective mode area of the cavity mode defined as

$$S_{\text{eff}} = \int_{z=0} U_E(x,y)\,dxdy, \quad (2)$$

where $U_E$ denotes the electric energy density of the cavity mode normalized to 1 at the cavity center. $\eta_{\text{far}}$ represents the far-field spatial matching between the incident light source and cavity mode, defined as

$$\eta_{\text{far}} = \left|\int_{|k_\parallel|<\frac{2\pi}{\lambda_0}} \boldsymbol{E}^*_{\text{cav}}(k_x, k_y) \cdot \boldsymbol{E}_{\text{source}}(k_x, k_y)\,dk_x dk_y\right|^2, \quad (3)$$

where $\boldsymbol{E}_{\text{cav}}$ and $\boldsymbol{E}_{\text{source}}$ are the field distributions of the cavity mode and the incident light source in the momentum space, respectively, obtained at the top surface of the slab [41]. $\boldsymbol{E}_{\text{cav}}$ is normalized within the light line of the cavity mode ($|k_\parallel| = \sqrt{k_x^2 + k_y^2} < 2\pi/\lambda_0$) such that the integral yields 1 when the far-field spatial matching of the excitation source to the cavity mode is satisfied. The last term in Eq. (1), $M$, represents a rate matching ($Q$ matching) between the absorption process and leakage of the cavity mode into the top direction and can be represented using the terms $\gamma_t$, $\gamma_b$, and $\gamma_{abs}$, which denote the cavity mode decay rates to the top, bottom, and owing to the QW absorption, respectively. We assumed that the decay rates in the lateral direction and due to metal absorption were minor (confirmed by separate 3D-FDTD simulations).

$$M = \frac{4\gamma_{abs}\gamma_t}{(\gamma_t + \gamma_b + \gamma_{abs})^2}. \quad (4)$$

We rewrite Eq. (4) in terms of $\Phi \equiv \gamma_t/\gamma_{rad}$, which describe the asymmetry in radiation direction, and total decay rate of the cavity mode due to the radiation, $\gamma_{rad} = \gamma_t + \gamma_b$, as follows:

$$M = \Phi \frac{4\gamma_{abs}\gamma_{rad}}{(\gamma_{abs} + \gamma_{rad})^2}, \quad (5)$$

or using corresponding quality factor representations:

$$M = \frac{Q_{rad}}{Q_t} \frac{4Q_{abs}Q_{rad}}{(Q_{abs} + Q_{rad})^2}. \quad (6)$$

$M$ takes the maximum value of 0.5 in vertically-symmetric cavities ($\gamma_t = \gamma_b$, that is, $\Phi = 0.5$) only when the absorption-radiation matching ($\gamma_{rad} = \gamma_{abs}$) is satisfied, which is the case discussed in our previous study [33]. In contrast, for an asymmetric cavity like the cavity shown in Fig. 1(a), $\Phi > 0.5$ can be achieved under some conditions, offering the possibility to exceed the efficiency limit in the symmetric case. In the following analysis, we used $Q_{abs} = 193$ [33] and $Q_{rad}$ was calculated by FDTD simulations using the absorption coefficient of GaAs QW layer of 0. $Q_{rad}/Q_t$ was obtained by calculating the fraction of the power radiated to the top direction to the total radiated power from the cavity.

Equations (1) and (6) provide useful information for absorption enhancement; cavity modes that are well localized, spatially matched to the excitation source, and satisfy the absorption-radiation matching ($Q_{rad} = Q_{abs}$) are beneficial. Furthermore, achieving a large $\Phi$ has a linear effect on $M$, indicating the importance of directional coupling to the top.

## 3. Result and discussion

Figure. 2 shows the dependence of $\eta_{abs}^{QD}(\lambda_0)$ on $T_{gap}$, calculated using Eq. (1). Each term in Eq. (1) was calculated using the corresponding equation in Sec. 2.2 with the field distributions and quality factors of the cavity mode obtained from the 3D-FDTD simulation. We also show the results calculated by independent 3D-FDTD simulations, which directly evaluate $\eta_{abs}^{QD}(\lambda)$. The results based on Eq. (1) and from direct calculations show a nice agreement, which demonstrates the validity of the analytical expression discussed in 2.2. $\eta_{abs}^{QD}(\lambda_0)$ oscillates with a period of approximately half of the resonant wavelength ($T_{gap}/\lambda_0 \sim 0.5$). Therefore, we intuitively understand that this phenomenon is related to the interference with the light reflected from the substrate. To analyze this phenomenon further, we calculated the approximate phase accumulation $\Delta\phi$ after a round trip in the air gap and half of the slab by using Eq. (7) [39]:

$$\Delta\phi = 2\left(\frac{2\pi}{\lambda_0}\right)T_{gap} + \psi + \left(\frac{2\pi}{\lambda_0}n_{slab}\right)T_{slab}, \quad (7)$$

where $\psi$ is the phase change after reflection and is supposed $\psi = \pi$. $\Delta\phi$ plotted as a function of $T_{gap}/\lambda_0$ represented by the red line in Fig. 2 shows that the in-phase (out-of-phase) conditions are almost satisfied at around $T_{gap}/\lambda_0 = 0.4$ ($T_{gap}/\lambda_0 = 0.65$). At $T_{gap}/\lambda_0 = 0.4$, where the in-phase condition is satisfied, $\eta_{abs}^{QD}(\lambda)$ is ~1.6 times higher than that for the cavity without substrate. Hereafter, we discuss each factor in Eq. (1) separately to understand the dominant factors affecting absorption enhancement.

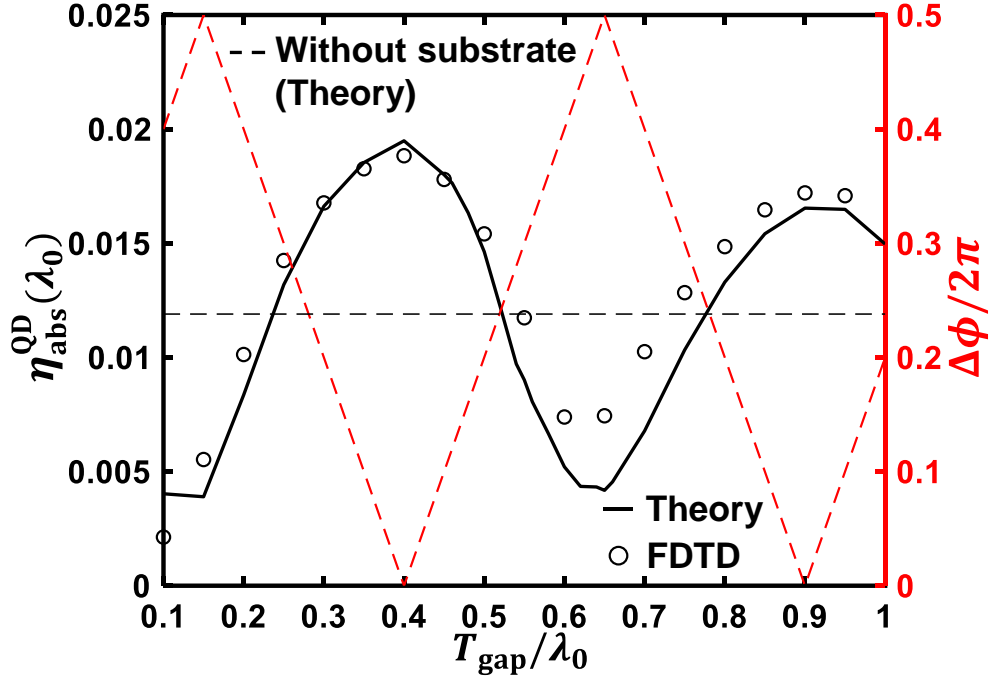

Fig. 2. Calculated absorption efficiency $\eta_{\text{abs}}^{\text{QD}}(\lambda_0)$ as a function of $T_{\text{gap}}$. Results obtained by Eq. (1) and individual 3D-FDTD simulations are represented by the black curve and circles, respectively. Accumulated phase after a round trip in the air gap and half of the slab calculated by Eq. (7) is plotted as the dashed line in the right $y$-axis.

Fig. 3 shows the dependence of $S_{\text{QD}}/S_{\text{eff}}$, $\eta_{\text{far}}$, $M$, and $\eta_{\text{abs}}^{\text{QD}}(\lambda_0)$ on $T_{\text{gap}}$. The corresponding values for the structure without considering the substrate (free-standing cavity) are shown as dashed lines for comparison. The change in $S_{\text{QD}}/S_{\text{eff}}$ was relatively small, which can be understood from the fact that the near-field distributions of the cavity modes change less when $T_{\text{gap}}$ changes. On the other hand, both $\eta_{\text{far}}$ and $M$ were significantly affected by the variation in $T_{\text{gap}}$. Given the fact that the trend of $\eta_{\text{abs}}^{\text{QD}}(\lambda_0)$ is almost identical to that of $M$, the improvement in $M$ is the primary important where the higher efficiency $\eta_{\text{abs}}^{\text{QD}}(\lambda_0)$ is obtained compared to the free-standing cavity. This interpretation will be also confirmed through a quantitative comparison shown later (see Table 1).

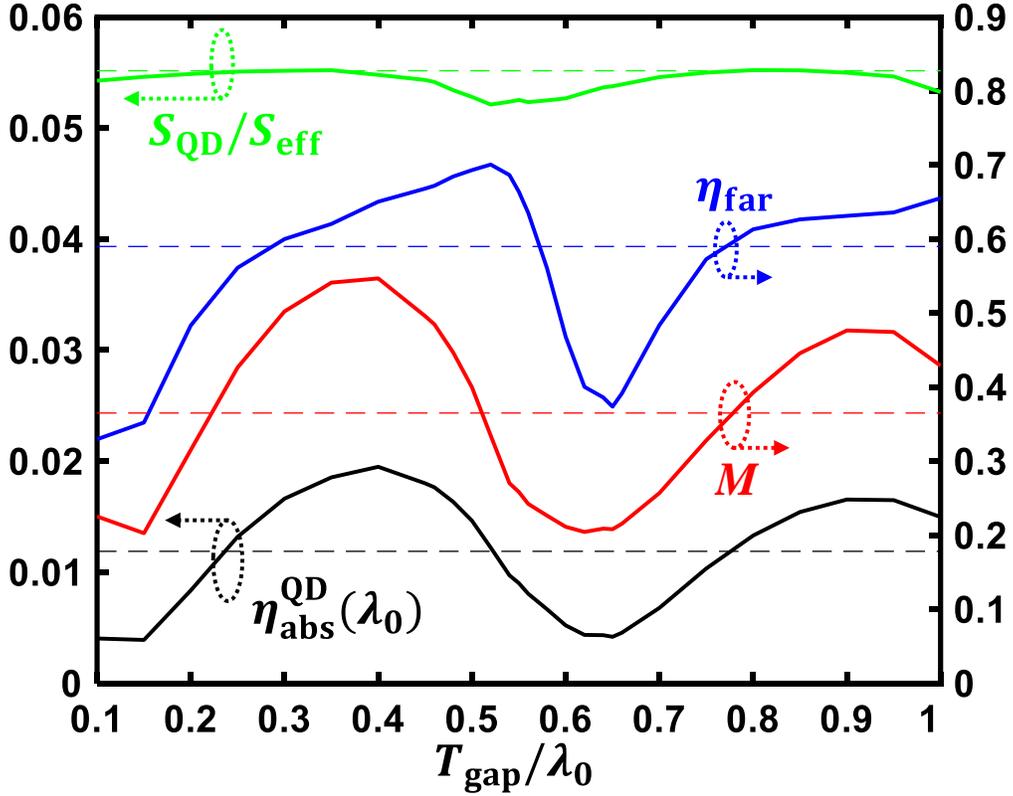

Fig. 3. Dependence of $M$, $\eta_{\text{far}}$, $S_{\text{QD}}/S_{\text{eff}}$, and $\eta_{\text{abs}}^{\text{QD}}(\lambda_0)$ on $T_{\text{gap}}$. Dashed lines represent the corresponding values for the structure without considering the substrate.

As shown in Eq. (5), $M$ is composed of $\Phi$ and the remaining part. The latter, $M/\Phi$, represents how perfectly the $Q$ matching condition is satisfied. To investigate which factor dominates the change in $M$, we plot the dependence of $M$, $\Phi$, and $M/\Phi$ on $T_{\text{gap}}$ in Fig. 4(a). Although both $\Phi$ and $M/\Phi$ change as $T_{\text{gap}}$ changes, $M$ roughly follows $\Phi$, meaning that the increase in asymmetry in the radiation leakage of the cavity mode plays the dominant role rather than the variation in $M/\Phi$ in the present cavity design. While $\Phi$ varies from ~0.2 to ~0.6, the variation in $M/\Phi$ is less significant. Although $Q_{\text{rad}}$ varies more than twice from 52 to125 (Fig. 4(a)), $M/\Phi$ changes only in the range from ~0.7 to ~0.95 with $Q_{\text{abs}} = 193$ (Fig. 4(b)). At $T_{\text{gap}}/\lambda_0 = 0.4$, where $M$ and $\eta_{\text{abs}}^{\text{QD}}(\lambda_0)$ show their maxi, the $Q$ matching condition is not fully satisfied. Further improvement in efficiency would be attainable if we could increase $Q_{\text{rad}}$ while keeping $\Phi$.

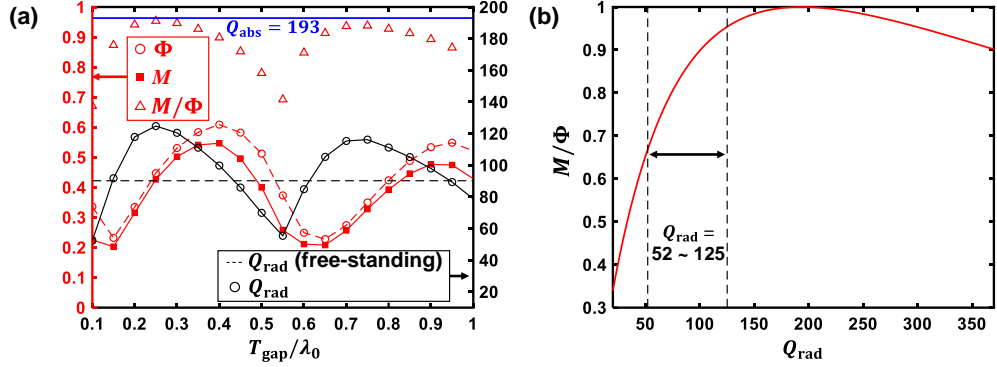

Fig. 4. (a) Dependence of $M$, $\Phi$, and $M/\Phi$ on $T_{\text{gap}}$. Variation of $Q_{\text{rad}}$ is also shown as black circles. $Q_{\text{rad}}$ for the free-standing cavity and $Q_{\text{abs}} = 193$ are shown as black dashed and blue lines, respectively. (b) Value of $M/\Phi = 4Q_{\text{abs}}Q_{\text{rad}}/(Q_{\text{abs}} + Q_{\text{rad}})^2$ when $Q_{\text{rad}}$ is varied from 20 to 370. The variation range of $Q_{\text{rad}}$ for the structures discussed are represented as dashed lines.

The dependence of $\eta_{\text{far}}$ on $T_{\text{gap}}$ is similar to that of $\Phi$, with a slight difference as it reaches a maximum at approximately $T_{\text{gap}}/\lambda_0 = 0.52$ instead of 0.4. The variation in $\eta_{\text{far}}$ (Fig. 5(a)) comes from the change in the overlap between the radiation pattern form the cavity mode toward the top direction and the beam pattern of the excitation Gaussian beam (Fig. 5(b)). In Figs. 5(c)-(f), we show the far-field intensity ($|\boldsymbol{E}|^2$) distributions for the structures with no substrate (the reference free-standing structure), $T_{\text{gap}}/\lambda_0 = 0.4$, 0.52, and 0.65, respectively. Each $|\boldsymbol{E}|^2$ was normalized such that the total field intensity integrated over the upper hemisphere was unity (Note that the color scales for Fig. 5(b)-(f) are the same). The far-field patterns for $T_{\text{gap}}/\lambda_0 = 0.4$ (Fig. 5(d)) and 0.52 (Fig. 5(e)) are more concentrated within the region of NA < 0.5 (corresponding to the half divergence angle < 30°), resulting in higher peak intensities than that of the free-standing case (Fig. 5(c)). In contrast, the far-field pattern of $T_{\text{gap}}/\lambda_0 = 0.65$ case distributes more uniformly over the larger divergence angle, decreasing $\eta_{\text{far}}$ significantly.

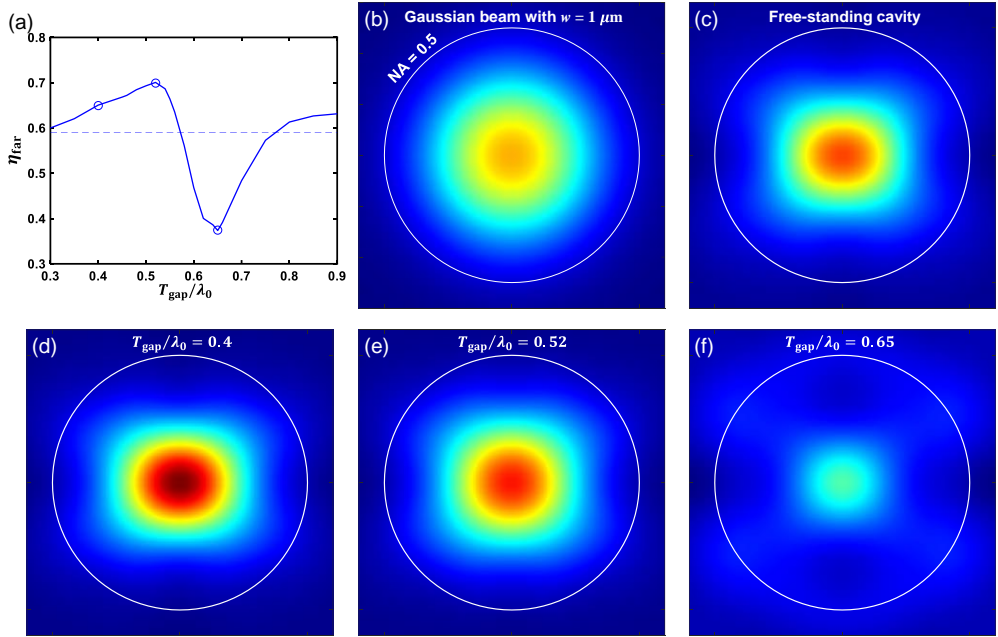

Fig. 5. (a) Dependence of $\eta_{\text{far}}$ on $T_{\text{gap}}$. Dashed line represents $\eta_{\text{far}}$ for free-standing cavity. Three marked spots ($T_{\text{gap}}/\lambda_0$ = 0.4, 0.5, and 0.65) represent $T_{\text{gap}}$ yielding the maximum $\eta_{\text{abs}}^{\text{QD}}(\lambda_0)$, maximum $\eta_{\text{far}}$, and minimum $\eta_{\text{far}}$, respectively. (b) Far-field intensity patterns of the Gaussian beam ($w$ = 1 μm), (c)-(f) Far-field intensity patterns of the $x$-polarized mode in the cavities with no substrate, $T_{\text{gap}}/\lambda_0$ = 0.4, 0.5, and 0.65, respectively.

Table 1. Comparison of $M$, $\eta_{\text{far}}$, $S_{\text{QD}}/S_{\text{eff}}$, and $\eta_{\text{abs}}^{\text{QD}}$ for bull's-eye cavities with/without considering the substrate.

|  | $T_{\text{gap}}$ | $M$ | $\eta_{\text{far}}$ | $S_{\text{QD}}/S_{\text{eff}}$ | $\eta_{\text{abs}}^{\text{QD}}(\lambda_0)$ |
|---|---|---|---|---|---|
| No substrate |  | 0.365 | 0.59 | 0.0552 | 0.0119 |
| With substrate | 328 nm ($0.4 \times \lambda_0$) | 0.547 | 0.651 | 0.0548 | 0.0195 |
|  | 533 nm ($0.65 \times \lambda_0$) | 0.208 | 0.374 | 0.0538 | 0.0042 |

Table 1 summarizes the values of $S_{\text{QD}}/S_{\text{eff}}$, $\eta_{\text{far}}$, and $M$ that contribute to the largest and smallest absorption efficiencies. The absorption efficiency at $T_{\text{gap}}/\lambda_0$ = 0.4, $\eta_{\text{abs}}^{\text{QD}}(\lambda_0)$ = 0.0195, is approximately 1.64 times enhanced than that of the free-standing situation ($\eta_{\text{abs}}^{\text{QD}}(\lambda_0)$ = 0.0119), where $M$ and $\eta_{\text{far}}$ show an improvement of around 50% and 10%, respectively, whereas the reduction in $S_{\text{QD}}/S_{\text{eff}}$ is less than 1%. These results confirm again that the improvement in $M$ dominates the improvement in absorption efficiency. At $T_{\text{gap}}/\lambda_0$ = 0.65, $\eta_{\text{abs}}^{\text{QD}}(\lambda_0)$ is 0.0042, which is ~0.35 times of $\eta_{\text{abs}}^{\text{QD}}(\lambda_0)$ for no substrate case. The reductions of $M$ and $\eta_{\text{far}}$ are responsible for the efficiency drop.

A proper cavity design can improve $M$ and $\eta_{\text{far}}$ to further improve the absorption efficiency. Achieving $Q_{\text{rad}}$ close to $Q_{\text{abs}}$ allows $M$ to be close to the maximally achievable $\Phi$. $\eta_{\text{far}}$ can be as large as 1 if the emission profile of the cavity mode is spatially matched to that of the excitation source. Field-based design methods [42,43] can be applied to precisely design the emission profile. For bull's-eye cavities, a simpler approach is to control the divergence of the emission profile by adjusting the size of the circular cavity. By increasing the cavity size, a narrower far-field emission pattern can be obtained. In addition, the increase in the quality factor due to the gentle spatial confinement in large-cavity bull's-eye structures can be beneficial for achieving a larger absorption-radiation matching $M/\Phi$. However, because a large cavity also causes a decrease in $S_{\text{QD}}/S_{\text{eff}}$, a proper optimization that considers the trade-off between $M \times \eta_{\text{far}}$ and $S_{\text{QD}}/S_{\text{eff}}$ is required in the future.

## 4. Conclusion

We numerically investigated the photon absorption efficiency of a gate-defined QD embedded in a bull's eye cavity with a substrate underneath. Our results demonstrated that the absorption efficiency can be improved more than 1.6 times compared to that without the substrate when the distance between the bull's-eye structure and the substrate is approximately 0.4 times the resonant wavelength. The substrate acted as a partial mirror that reflected light, thereby modifying the vertical coupling constant of the cavity. At the distance, the constructive interference condition for increasing upward coupling is satisfied, which resulted in a 1.64 times improvement of the absorption efficiency. In experiments, the distance between the bull's-eye structure and the substrate can be controlled precisely by adjusting the thickness of the sacrificial layer.

The analysis of each factor determining the absorption efficiency clarified that the upward directionality, $\Phi = \gamma_t/\gamma_{\text{rad}}$, plays a dominant role in the improvement. We also found that the far-field spatial overlap between the cavity mode and the incident Gaussian beam, $\eta_{\text{far}}$, is also favored by the presence of substrate, suggesting that further improvement is possible by tailoring the cavity radiation profile through careful tuning of the cavity design in the future.

The substrate-induced asymmetry can improve absorption in a simple manner. This simple method would be useful for improving the collection efficiency of photons emitted from a gate-defined QD [44]. We would also emphasize that this substrate effect is applicable to other host materials of solid-state qubits, such as bulk single-crystal diamonds [45].

**Funding.** JST CREST (JPMJCR15N2); MEXT KAKENHI (22H04962); JST Moonshot R&D (JPMJMS2066).
**Acknowledgements.** The authors thank Yasutomo Ota for fruitful discussions.
**Disclosures.** The authors declare no conflicts of interest.
**Data availability.** Data underlying the results presented in this paper are not publicly available at this time but may be obtained from the authors upon reasonable request.